\documentclass{llncs}

\usepackage{cite}
\usepackage{amsmath}
\usepackage{mathtools}
\usepackage{amssymb}
\usepackage{amsfonts}
\let\proof\relax

\usepackage{amsthm}
\usepackage[linesnumbered]{algorithm2e}
\usepackage{graphicx}
\usepackage{textcomp}
\usepackage{enumitem}
\usepackage{url}
\usepackage{floatrow}
\usepackage{wrapfig}
\usepackage{algorithmic}
\usepackage{subcaption}

\usepackage[printwatermark]{xwatermark}

\DeclarePairedDelimiter\floor{\lfloor}{\rfloor}

\usepackage{tikz}
\usepackage{tikz-3dplot}

\newcommand{\Comment}[1]{}
\newtheorem{Definition}{Definition}

\newtheorem{Proposition}{Proposition}

\theoremstyle{remark}
\newcommand{\sig}[1]{\textsf{{#1}\xspace}}
\newcommand{\countappear}[1]{\{#1\}}

\newfloatcommand{capbtabbox}{table}[][\FBwidth]

\begin{document}

\title{Quantitative Projection Coverage for  Testing ML-enabled Autonomous Systems
}

\author{
Chih-Hong Cheng\inst{1}
\and
Chung-Hao Huang\inst{1}
\and
Hirotoshi Yasuoka\inst{2}
}

\institute{
	fortiss - Landesforschungsinstitut des Freistaats Bayern \\
	\and
	\vspace{1mm}
	DENSO CORPORATION \\
    \vspace{1mm}
    \texttt{\{cheng,huang\}@fortiss.org, hirotoshi\_yasuoka@denso.co.jp}\\
}

\maketitle

\begin{abstract}

Systematically testing models learned from neural networks remains a crucial unsolved barrier to successfully justify safety for autonomous vehicles engineered using data-driven approach. 
We propose quantitative $k$-projection coverage as a metric to mediate combinatorial explosion while guiding the data sampling process. 
By assuming that domain experts propose largely independent environment conditions and by associating elements in each condition with weights, the product of these conditions forms scenarios, and one may interpret weights associated with each equivalence class as relative importance.
Achieving full $k$-projection coverage requires that the data set, when being projected to the hyperplane formed by arbitrarily selected $k$-conditions, covers each class with number of data points no less than the associated weight.  For the general case where scenario composition is \emph{constrained} by rules, precisely computing $k$-projection coverage remains in~\sig{NP}. In terms of finding minimum test cases to achieve full coverage, 
we present theoretical complexity for important sub-cases and an encoding to 0-1 integer programming.  We have implemented a 
research prototype that generates test cases for a visual object detection unit in automated driving, demonstrating the technological feasibility of our proposed coverage criterion.

\end{abstract}

\section{Introduction}

 There is a recent hype of applying neural networks in automated driving, ranging from perception~\cite{chen2015deepdriving,huval2015empirical} to the creation of driving strategies~\cite{Lenz2017,sun2017fast} to even end-to-end driving setup~\cite{DBLP:journals/corr/BojarskiTDFFGJM16}. Despite many public stories that seemly hint the technical feasibility of using neural networks, one fundamental challenge is to 
 establish rigorous safety claims by considering all classes of relevant scenarios whose presence is subject to technical or societal constraints.

 The key motivation of this work is that, 
 apart from recent formal verification efforts~\cite{DBLP:conf/cav/HuangKWW17,ehlers2017formal,DBLP:conf/cav/KatzBDJK17,cheng2017maximum} where scalability and lack of specification are obvious concerns, the most plausible approach, from a certification perspective, remains to be testing.  
 As domain experts or authorities in autonomous driving may suggest~$n$ (incomplete) weighted criteria for describing the operating conditions such as weather, landscape, or partially occluding pedestrians, with these criteria one can systematically partition the domain and weight each partitioned class based on its relative importance. This step fits very well to the consideration as in automotive safety standard ISO 26262, where for deriving test cases, it is highly recommended to \emph{perform analysis of equivalence classes} (Chap 6, Table 11, item 1b). Unfortunately, there is an \emph{exponential} number of classes being partitioned, making the na\"{\i}ve coverage metric of having at least one data point in each class unfeasible. In addition, such a basic metric is \emph{qualitative} in that it does not address the relative importance among different scenarios.

Towards above issues, in this paper we study the problem of \emph{quantitative $k$-projection coverage}, i.e., for arbitrary $k$ criteria being selected ($k \ll n$ being a small \emph{constant} value), the data set, when being projected onto the $k$-hyperplane, needs to have  (in each region) data points no less than the associated weight. When~$k$ is a constant, the size of required data points to achieve full quantitative $k$-projection coverage remains polynomially bounded. Even more importantly, for the case where the composition of scenarios is \emph{constrained} by rules,
we present an $\sig{NP}$ algorithm to compute exact $k$-projection coverage. This is in contrast to the case without projection, where computing exact coverage is~$\sharp \sig{P}$-hard.   

Apart from calculating coverage, another crucial problem is to generate, based on the goal of increasing coverage, fewer scenarios if possible, as generating images or videos matching the scenario in autonomous driving is largely semi-automatic and requires huge human efforts. While we demonstrate that for \emph{unconstrained} quantitative $1$-projection, finding a minimum set of test scenarios to achieve full coverage remains in polynomial time, we prove that for $3$-projection, the problem is \sig{NP}-complete. To this end, we develop an efficient encoding to 0-1 integer programming which allows incrementally creating scenarios to maximally increase coverage.

To validate our approach, we have implemented a prototype to define and ensure coverage of a vision-based front-car detector. The prototype has integrated state-of-the-art traffic simulators and image synthesis frameworks~\cite{mirza2014conditional,wang2017high}, in order to synthesize close-to-reality images specific to automatically proposed scenarios.

\tdplotsetmaincoords{60}{125}

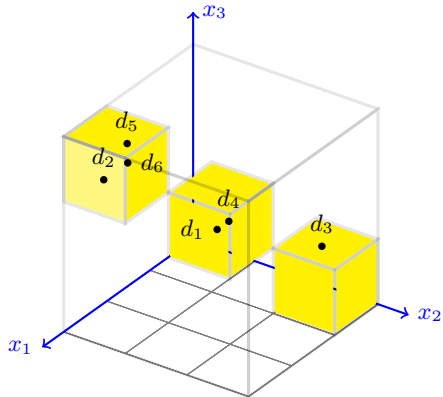
\begin{figure}[t]
	\centering
	\begin{tikzpicture}
	[tdplot_main_coords,
	grid/.style={very thin,gray},
	axis/.style={->,blue,thick},
	cube/.style={opacity=.1,very thick}]
	\foreach \x in {0,1,...,3}
	\foreach \y in {0,1,...,3}
	{
		\draw[grid] (\x,0) -- (\x,3);
		\draw[grid] (0,\y) -- (3,\y);
	}			
	
	\draw[axis] (0,0,0) -- (3.5,0,0) node[anchor=east]{$x_1$ };
	\draw[axis] (0,0,0) -- (0,3.5,0) node[anchor=west]{$x_2$ };
	\draw[axis] (0,0,0) -- (0,0,3.5) node[anchor=west]{$x_3$ };

	
	\draw[cube, fill=yellow, fill opacity=0.5] (3,0,2) -- (3,1,2) -- (3,1,3) -- (3,0,3) -- cycle;
	
	\draw[cube, fill=yellow, fill opacity=1] (2,1,2) -- (3,1,2) -- (3,1,3) -- (2,1,3) -- cycle;
	
	\draw[cube, fill=yellow, fill opacity=1] (2,0,3) -- (2,1,3) -- (3,1,3) -- (3,0,3) -- cycle;
	
	\draw[cube, fill=yellow, fill opacity=1] (2,1,1) -- (2,2,1) -- (2,2,2) -- (2,1,2) -- cycle;
	
	\draw[cube, fill=yellow, fill opacity=1] (1,2,1) -- (2,2,1) -- (2,2,2) -- (1,2,2) -- cycle;
	
	\draw[cube, fill=yellow, fill opacity=1] (1,1,2) -- (1,2,2) -- (2,2,2) -- (2,1,2) -- cycle;

	\draw[cube, fill=yellow, fill opacity=1] (1,2,0) -- (1,3,0) -- (1,3,1) -- (1,2,1) -- cycle;
	
	\draw[cube, fill=yellow, fill opacity=1] (0,3,0) -- (1,3,0) -- (1,3,1) -- (0,3,1) -- cycle;
	
	\draw[cube, fill=yellow, fill opacity=1] (0,2,1) -- (0,3,1) -- (1,3,1) -- (1,2,1) -- cycle;

	\draw[cube] (3,0,0) -- (3,3,0) -- (3,3,3) -- (3,0,3) -- cycle;
	
	\draw[cube] (0,3,0) -- (3,3,0) -- (3,3,3) -- (0,3,3) -- cycle;
	
	\draw[cube] (0,0,3) -- (0,3,3) -- (3,3,3) -- (3,0,3) -- cycle;
	
	\node[draw=none,shape=circle, fill, inner sep=1pt, label=above:{$d_2$}] (d2) at (2.5,0.3,2.2){};  
	
	\node[draw=none,shape=circle, fill, inner sep=1pt,label=above:{$d_5$}] (d5) at (2.1,0.4,2.6){};  
	
	\node[draw=none,shape=circle, fill, inner sep=1pt, label=right:{$d_6$}] (d6) at (2.8,0.9,2.8){};  

	\node[draw=none,shape=circle, fill, inner sep=1pt, label=left:{$d_1$}] (d1) at (1.3,1.3,1.2){};  
	
	\node[draw=none,shape=circle, fill, inner sep=1pt, label=above:{$d_4$}] (d4) at (1.6,1.7,1.6){};  
	
	\node[draw=none,shape=circle, fill, inner sep=1pt,label=above:{$d_3$}] (d3) at (0.3,2.3,0.8){};  

	\end{tikzpicture}
	
	\caption{A total of $6$ data points and their corresponding equivalence classes (highlighted as bounding boxes).}
	\label{fig.sample}
	\vspace{-3mm}
\end{figure}

 \vspace{3mm}
\noindent \textbf{(Related Work)} The use of AI technologies, in particular the use of neural networks, has created fundamental challenges in safety certification. 
Since 2017 there has been a tremendous research advance in formally verifying properties of neural networks, with focuses on neurons using piecewise linear activation function (ReLU). For sound-and-complete approaches, Reluplex and Planet developed specialized rules for managing the 0-1 activation in the proof system~\cite{DBLP:conf/cav/KatzBDJK17,ehlers2017formal}. Our previous work~\cite{cheng2017maximum,cheng2018neural} focused on the reduction to mixed integer liner programming (MILP) and applied techniques to compute tighter bounds such that in MILP, the relaxation bound is closer to the real bound. Exact approaches suffer from combinatorial explosion and currently the verification speed is not satisfactory. For imprecise yet sound approaches, recent work has been emphasizing linear relaxation of ReLU units by approximating them using outer convex polytopes~\cite{weng2018towards,kolter2017provable,sinha2017certifiable}, making the verification problem feasible for linear programming solvers. These approaches are even applied in the synthesis (training) process, such that one can derive provable guarantees~\cite{kolter2017provable,sinha2017certifiable}. Almost all verification work (apart from~\cite{DBLP:conf/cav/KatzBDJK17,ehlers2017formal,cheng2018neural}) targets robustness properties, which is similar to adversarial testing (e.g., FGSM \& iterative attacks~\cite{szegedy2013intriguing}, deepfool~\cite{moosavi2016deepfool}, Carlini-Wagner attacks~\cite{carlini2017towards}) as in the machine learning community. All these approaches can be complemented with our approach by having our approach  covering  important scenarios, while adversarial training or formal verification measuring robustness within each scenario.

For classical structural coverage testing criteria such as MC/DC, they fail to deliver assurance promises,  as satisfying full coverage either turns trivial ($\sig{tanh}$) or intractable (ReLU). 
The recent work by Sun, Huang, and Kroening~\cite{sun2018testing} borrows the concept of MC/DC and considers a structural coverage criterion, where one needs to find tests to ensure that for every neuron, its activation is supported by independent activation of neurons in its immediate previous layer.
Such an approach can further be supported by concolic testing, as being recently demonstrated by same team~\cite{sun2018concolic}. Our work and theirs should be viewed as complementary, as we focus on the data space for training and testing neural networks, while they focus on the internal structure of a neural network. However, as in the original MC/DC concept, each condition in a conditional statement (apart from detecting errors in programming such as array out-of-bound which is not the core problem of neural networks) is designed to describe scenarios which should be viewed as natural consequences of input space partitioning (our work). Working on coverage criteria related to the internal structure of neural networks, provided that one cannot enforce the meaning of an individual neuron but  can only empirically analyze it via reverse engineering (as in standard approaches like saliency maps~\cite{simonyan2013deep}), is less likely provide direct benefits. Lastly, one major benefit of these structural testing approaches, based on the author claims, is to find adversarial examples via perturbation, but the benefit may be reduced due to  new training methods with provable perturbation bounds~\cite{kolter2017provable,sinha2017certifiable}.

Lastly, our proposed metric tightly connects to the classic work of combinatorial testing and covering arrays~\cite{colbourn2004combinatorial,nie2011survey,lawrence2011survey,lei1998parameter,seroussi1988vector}. However, as their application starts within hardware testing (i.e., each input variable being \sig{true} or \sig{false}), the quantitative aspects are not really needed and it does not need to consider constrained input cases, which is contrary to our practical motivation in the context of autonomous driving. For unconstrained cases, there are some results of \sig{NP}-completeness in the field of combinatorial testing, which is largely based on the proof in \cite{seroussi1988vector}. It is not applicable to our case, as the proof is based on having freedom to define the set of groups to be listed in the projection. In fact, as listed in a survey paper~\cite{lawrence2011survey}, the authors commented that it remains open whether ``the problem of generating a minimum test set for pairwise
testing ($k=2$) is \sig{NP}-complete" and ``existing proof in~\cite{lei1998parameter} for the \sig{NP}-completeness of pairwise testing is wrong" (due to the same reason where pairwise testing cannot have freedom to define the set of groups). Our new \sig{NP}-completeness result  in this paper can be viewed as a relaxed case by considering $k=3$ with sampling being quantitative than qualitative.

\section{Discrete Categorization and Coverage}

Let $\mathcal{DS} \subset \mathbb{R}^m$ be the \emph{data space},  $\mathcal{D} \subset \mathcal{DS}$ be a finite set called \emph{data set}, and $d\in \mathcal{DS}$ is called a \emph{data point}. 
A \emph{categorization} $\mathcal{C} = \langle C_1, \ldots, C_n \rangle$ is a list of functions that transform any data point $d$ to a \emph{discrete categorization point}~$\mathcal{C}(d) = (C_1(d), \ldots, C_n(d))$, where for all $i \in \{1, \ldots, n\}$, $C_i$ has co-domain $\{\sig{0}, \sig{1}, \ldots , \alpha\}$. 
Two data points $d_1$ and $d_2$ are \emph{equivalent by categorization}, denoted by $d_1 \equiv_{\mathcal{C}} d_2$, if $\mathcal{C}(d_1) = \mathcal{C}(d_2)$. The \emph{weight} of a categorization $\mathcal{W} = \langle W_1, \ldots, W_n \rangle$ further assigns value $j \in \{\sig{0}, \sig{1}, \ldots , \alpha\}$ in the co-domain of $C_i$ with an integer value $W_i(j) \in \{0 ,\ldots, \beta\}$.

\vspace{2mm}
Next, we define constraints over categorization, allowing domain experts to express knowledge by specifying relations among categorizations. Importantly, for all data points in the data space, whenever they are transformed using $\mathcal{C}$, the transformed discrete categorization points satisfy the constraints.  

\begin{Definition}[Categorization constraint]
 A categorization constraint $\mathcal{CS} = \{CS_1, \ldots, CS_p\}$ is a set of constraints with each being a CNF formula having literals of the form $C_i(d) \;\emph{\sig{op}}\; \alpha_i$, where  $\emph{\sig{op}} \in \{=, \neq\}$ and $\alpha_i \in \{\emph{\sig{0}}, \ldots, \alpha\}$. 
\end{Definition}

Let $\odot_{i\in \{1,\ldots,n\}} W_i(c_i)$ abbreviate  $W_1(c_1)\;\odot \ldots \odot W_n(c_n)$, where $\odot \in \{+, \times , \sig{max}\}$ can be either scalar addition, multiplication, or max operators. In this paper, unless specially mentioned we always treat $\odot$ as scalar multiplication. Let $\mathcal{C}(\mathcal{D})$ be the multi-set $\{\mathcal{C}(d) \;|\; d\in \mathcal{D}\}$, and $\leq^\mathcal{W}_{\odot}$ be set removal operation on $\mathcal{C}(\mathcal{D})$ such that every categorization point $(c_1, \ldots, c_n) \in \mathcal{C}(\mathcal{D})$ has at most cardinality equal to $\odot_{i\in \{1,\ldots,n\}} W_i(c_i)$.
We define \emph{categorization coverage} by requiring that for each discrete categorization point $(c_1, \ldots, c_n)$, in order to achieve full coverage,  have at least $\odot_{i\in \{1,\ldots,n\}} W_i(c_i)$ data points.

\begin{Definition}[Categorization coverage]~\label{def.categorization.coverage}
Given a data set~$\mathcal{D}$, a categorization~$\mathcal{C}$ and its associated weights $\mathcal{W}$, define the categorization coverage $\emph{\sig{cov}}_{\mathcal{C}}(\mathcal{D})$ for data set $\mathcal{D}$ over $\mathcal{C}$ and $\mathcal{W}$ to be $\frac{|\leq^\mathcal{W}_{\odot}(\mathcal{C}(\mathcal{D}))|}{ \sum_{(c_1, \ldots, c_n)\in \emph{\sig{sat}}(\mathcal{CS)}} \odot_{i\in \{1,\ldots,n\}} W_i(c_i) }$, where  $\emph{\sig{sat}} (\mathcal{CS})$ is the set of discrete categorization points satisfying constraints $\mathcal{CS}$.
\end{Definition}

\noindent \textbf{(Example 1)} In Fig.~\ref{fig.sample}, let the data space $\mathcal{DS}$ be $[0, 3)\times[0, 3)\times[0, 3)$ and the data set be $\mathcal{D} = \{d_1, \ldots, d_6\}$. By setting $\mathcal{C} = \langle C_1, C_2, C_3\rangle$ where $C_i = \floor*{x_i}$ for $i \in \{1,2,3\}$, then for data points $d_2$, $d_5$ and $d_6$, applying $C_1, C_2$ and $C_3$ creates  $\mathcal{C}(d_2)=\mathcal{C}(d_5)=\mathcal{C}(d_6)= (\sig{2}, \sig{0}, \sig{2})$, i.e., $d_2 \equiv_{\mathcal{C}} d_5 \equiv_{\mathcal{C}} d_6$. Similarly, $d_1 \equiv_{\mathcal{C}} d_4$.


\begin{itemize}
	\item If $\mathcal{CS}$ is an empty set and $\forall i\in \{1, 2, 3\}, j\in \{\sig{0}, \sig{1}, \sig{2}\}: W_i(j) = 1$, then $|\sig{sat}(\mathcal{CS})|=3^3=27$, $\leq^\mathcal{W}_{\odot}(\mathcal{C}(\mathcal{D}))$ removes $\mathcal{C}(d_2),\mathcal{C}(d_4),\mathcal{C}(d_5)$ by keeping one element in each equivalence class, and  $\sig{cov}_{\mathcal{C}}(\mathcal{D})$ equals~$\frac{3}{27} = \frac{1}{9}$.
	
	\item If $\mathcal{CS} = \{(C_1(d) \neq \sig{0} \;\vee\; C_2(d) = \sig{2})\}$ and $\forall i\in \{1, 2, 3\}, j\in \{\sig{0}, \sig{1}, \sig{2}\}: W_i(j) = 1$, then $|\sig{sat}(\mathcal{CS})| = 21$ rather than~$27$ in the unconstrained case, and  $\sig{cov}_{\mathcal{C}}(\mathcal{D})$ equals~$\frac{3}{21} = \frac{1}{7}$. Notice that all data points, once when being transformed into discrete categorization points, satisfy the categorization constraint.
	
	\item Assume that $\mathcal{CS}$ is an empty set, and $\forall i\in \{1, 2, 3\}, j\in \{\sig{0}, \sig{1}, \sig{2}\}$,  $W_i(j)$  always returns~$1$ apart from $W_1(\sig{2})$ and $W_3(\sig{2})$ returning $3$.  Lastly, let $\odot$ be scalar multiplication. Then for discrete categorization points having the form of $(\sig{2},\sig{-},\sig{2})$, a total of $W_1(\sig{2}) \times W_3(\sig{2}) = 9$ data points are needed. One follows the definition and computes $\sig{cov}_{\mathcal{C}}(\mathcal{D})$ to be~$\frac{3+1+1}{9\times 3+3\times 12+1\times12} = \frac{1}{13}$.
\end{itemize}


\noindent Achieving $100\%$ categorization coverage is essentially hard, due to the need of exponentially many  data points.

\begin{Proposition}
	Provided that $\mathcal{CS} = \emptyset$ and $\forall i\in \{1, \ldots, n\}, j\in \{\sig{0}, \ldots, \alpha\}: W_i(j) = 1$, 
	to achieve full coverage where $\sig{cov}_{\mathcal{C}}(\mathcal{D}) = 1$, $|\mathcal{D}|$ is  exponential to the number of categorizations.

\end{Proposition}

\proof

	 Based on the given condition, $|\sig{sat}(\mathcal{CS})|= (\alpha+1)^{n}$, and for each $(c_i, \ldots, c_n) \in \sig{sat}(\mathcal{CS})$, $\odot_{i\in \{1,\ldots,n\}} W_i(c_i)=1$.  Therefore, $\sig{cov}_{\mathcal{C}}(\mathcal{D}) = \frac{|\leq^\mathcal{W}_{\odot}(\mathcal{C}(\mathcal{D}))|}{(\alpha+1)^{n}}$. As \\ $|\leq^\mathcal{W}_{\odot}(\mathcal{C}(\mathcal{D}))| \leq |\mathcal{C}(\mathcal{D})|$, to achieve full coverage  $|\mathcal{C}(\mathcal{D})|$ (and correspondingly~$|\mathcal{D}|$) needs to be exponential to the number of categorizations.
\qed

\begin{Proposition}
Computing exact $\sig{cov}_{\mathcal{C}}(\mathcal{D})$ is $\sharp\sig{P}$-hard.
\end{Proposition}

\proof Computing the exact number of the denominator in Definition~\ref{def.categorization.coverage}, under the condition of $\alpha = \sig{1}$, equals to the problem of model counting for a SAT formula, which is known to be $\sharp\sig{P}$-complete.
\qed

\section{Quantitative Projection Coverage}

The intuition behind quantitative projection-based coverage is that, although it is unfeasible to cover all discrete categorization points, one may degrade the confidence by asking if the data set has covered every pair or triple of possible categorization with sufficient amount of data. 


\begin{Definition}[$k$-projection]
	Let set $\Delta = \{\Delta_1, \ldots, \Delta_k\} \subseteq \{1, \ldots, n\}$ where elements in $\Delta$ do not overlap. Given $d\in\mathcal{D}$, define the projection of a discrete categorization point $\mathcal{C}(d)$ over $\Delta$ to be $\sig{Proj}_{\Delta}(\mathcal{C}(d)) = (C_{\Delta_1}(d), C_{\Delta_2}(d), \ldots, C_{\Delta_k}(d))$.
	
\end{Definition}

 Given a multi-set~$S$ of discrete categorization points, we use $\sig{Proj}_{\Delta}(S)$ to denote the resulting multi-set by applying the projection function on each element in~$S$, and analogously define $\leq^\mathcal{W}_{\odot_{\Delta}}(\sig{Proj}_{\Delta}(S))$ to be a function which removes elements in $\sig{Proj}_{\Delta}(S)$ such that every element $(c_{\Delta_1}, \ldots, c_{\Delta_k})$  has cardinality at most  $W_{\Delta_1}(c_{\Delta_1}) \odot \ldots \odot W_{\Delta_k}(c_{\Delta_k})$. 
Finally, we define $k$-projection coverage based on applying projection operation on the data set~$\mathcal{D}$, for all possible subsets of~$\mathcal{C}$ of  size $k$.

\begin{Definition}[$k$-projection coverage]~\label{def.k.projection}
	Given a data set $\mathcal{D}$ and categorization~$\mathcal{C}$, define the $k$-projection categorization coverage $\emph{\sig{cov}}^{k}_{\mathcal{C}}(\mathcal{D})$ for data set $\mathcal{D}$ over $\mathcal{C}$ and~$\mathcal{W}$ to be 
	
	\vspace{-2mm}
	\[
	\frac{  \sum_{\{\Delta \;:\; |\Delta| =k\}} |\leq^\mathcal{W}_{\odot_{\Delta}}(
		\sig{Proj}_{\Delta}(\mathcal{C}(\mathcal{D})
		)|}{  \sum_{\{\Delta \;:\; |\Delta| =k\}}  \sum_{(c_{\Delta_1}, \ldots, c_{\Delta_k})\in \emph{\sig{to-set}}(\sig{Proj}_{\Delta}(\emph{\sig{sat}}(\mathcal{CS)))}}
		W_{\Delta_1}(c_{\Delta_1})\odot \ldots \odot W_{\Delta_k}(c_{\Delta_k})}
	\] where function $\emph{\sig{to-set}}()$ translates a multi-set to a set without element repetition.

\end{Definition}

\noindent \textbf{(Example)} Consider again  Fig.~\ref{fig.sample} with $\odot$ being scalar multiplication, $\mathcal{CS} = \emptyset$ and $\forall i\in \{1, 2, 3\}, j\in \{\sig{0}, \sig{1}, \sig{2}\}: W_i(j) = 2$.
	\begin{itemize}
		\item  For $k=1$, one computes $\sig{cov}^{1}_{\mathcal{C}}(\mathcal{D}) = \frac{5+5+5}{{3\choose 1} 3^1 2^1} = \frac{15}{18}$. 
		In the denominator, $\Delta$ has ${3 \choose 1}$ choices, namely $\Delta = \{1\}$, $\Delta = \{2\}$, or $\Delta = \{3\}$. Here we do detailed analysis over $\Delta = \{1\}$, i.e., we consider the projection to  $C_1$.
		\begin{itemize}
			\item Since $\mathcal{CS} = \emptyset$ , $\sig{sat}(\mathcal{CS})$ allows all possible~$3^3$ assignments.
			\item $\sig{Proj}_{\Delta}(\sig{sat}(\mathcal{CS}))$ creates a set with elements $\sig{0}$, $\sig{1}$, $\sig{2}$ with each being repeated~$9$ times, and $\sig{to-set}(\sig{Proj}_{\Delta}(\sig{sat}(\mathcal{CS})))$ removes multiplicity and creates $\{\sig{0}, \sig{1}, \sig{2}\}$. The sum equals $W_1(\sig{0})+W_1(\sig{1})+W_1(\sig{2})=6$.
		\end{itemize}

		The ``$5$" in the numerator comes from the contribution of $(\sig{2,0,2})$ with~$2$ (albeit it has $3$ data points), $(\sig{1,1,1})$ with~$2$, and $(\sig{0,2,0})$ with~$1$.  
		\item  For $k=2$, one computes $\sig{cov}^{2}_{\mathcal{C}}(\mathcal{D}) = \frac{6+6+6}{{3\choose 2} 3^{2}2^2} = \frac{1}{6}$. The denominator captures three hyper planes ($x_1x_2$, $x_1x_3$, $x_2x_3$) with each having $3^2$ grids and with each grid allowing $2^2$ data points. 
	\end{itemize}	

\noindent Notice that Definition~\ref{def.categorization.coverage} and~\ref{def.k.projection} are the same when one takes~$k$ with value~$n$.

\begin{Proposition}
	$\sig{cov}^{n}_{\mathcal{C}}(\mathcal{D}) = \sig{cov}_{\mathcal{C}}(\mathcal{D})$.
\end{Proposition}

\proof  When $k=n$, the projection operator does not change $\sig{sat}(\mathcal{CS})$. Subsequently, $\sig{to-set}$ operator is not effective as $\sig{Proj}_{\Delta}(\sig{sat}(\mathcal{CS})) = \sig{sat}(\mathcal{CS})$ is already a set, not a multi-set. Finally, we also have $W_{\Delta_1}(c_{\Delta_1})\odot \ldots \odot W_{\Delta_k}(c_{\Delta_k}) = \odot_{i\in \{1,\ldots,n\}} W_i(c_i)$. Thus the denominator part of Definition~\ref{def.categorization.coverage} and~\ref{def.k.projection} are computing the same value. The argument also holds for the numerator part. Thus the definition of $	\sig{cov}^{n}_{\mathcal{C}}(\mathcal{D})$ can be rewritten as $\sig{cov}_{\mathcal{C}}(\mathcal{D})$. 
\qed 

\vspace{2mm}
The important difference between categorization coverage and $k$-projection  coverage (where $k$ is a constant) includes the number of data points needed to achieve full coverage (exponential vs. polynomial), as well as the required time to compute exact coverage ($\sharp\sig{P}$ vs. $\sig{NP}$).

\begin{Proposition}
If $k$ is a constant, then to satisfy full $k$-projection coverage, one can find a data set $\mathcal{D}$ whose size is bounded by a number which is polynomial to~$n$, $\alpha$ and $\beta$. 
\end{Proposition}

\proof In Definition~\ref{def.k.projection}, the denominator is bounded by ${n \choose k}(\alpha+1)^k\beta^k$. 
	\begin{itemize}
		\item  The total number of possible $\Delta$ with size $k$ equals~${n \choose k}$, which is a polynomial of~$n$ with highest degree being~$k$. 
		
		\item For each $\Delta$, $(c_{\Delta_1}, \ldots, c_{\Delta_k})\in \sig{to-set}(\sig{Proj}_{\Delta}(\sig{sat}(\mathcal{CS})))$ has at most $(\alpha+1)^k$ possible assignments - this happens when $\mathcal{CS}=\emptyset$.
		
		\item For each assignment of $(c_{\Delta_1}, \ldots, c_{\Delta_k})$, $W_{\Delta_1}(c_{\Delta_1}) \odot \ldots \odot W_{\Delta_k}(c_{\Delta_k})$ can at most has largest value $\beta^k$.
	\end{itemize}

As one can use one data point for each element in the denominator, $\mathcal{D}$ which achieves full coverage is polynomially bounded. 
\qed

\vspace{2mm}
\noindent \textbf{(Example 2)} Consider a setup of defining traffic scenarios where one has $\alpha = 3$ and $n=20$. When $\mathcal{CS}=\emptyset$ and $\forall i\in \{1, \ldots, n\}, j\in \{\sig{0}, \ldots, \alpha\}: W_i(j) = 1$, the denominator of categorization coverage as defined in Definition~\ref{def.categorization.coverage} equals $3486784401$, while the denominator of $2$-projection coverage equals~$1710$ and the denominator of $3$-projection coverage equals~$10260$.

\begin{Proposition}
	If $k$ is a constant, then computing $k$-projection coverage can be done in $\sig{NP}$. If $\mathcal{CS} = \emptyset$, then computing $k$-projection coverage can be done in $\sig{P}$.
\end{Proposition}

\proof 

\begin{itemize}
	\item For the general case where $\mathcal{CS} \neq \emptyset$, to compute $k$-projection coverage, the crucial problem is to know the precise value of the denominator. In the denominator, the part 
	``$(c_{\Delta_1}, \ldots, c_{\Delta_k})\in \sig{to-set}(\sig{Proj}_{\Delta}(\sig{sat}(\mathcal{CS)))}$'' is actually only checking if  
	for grid $(c_{\Delta_1}, \ldots, c_{\Delta_k})$ in the projected $k$-hyperplane, whether it is possible to be occupied due to the constraint of $\mathcal{CS}$.  If one knows that it can be occupied, simply add to the denominator by  $W_{\Delta_1}(c_{\Delta_1}) \odot \ldots \odot W_{\Delta_k}(c_{\Delta_k})$.
	This ``occupation checking'' step can be  achieved by examining the satisfiability of $\mathcal{CS}$ with $C_{\Delta_i}$ being replaced by the concrete assignment $(c_{\Delta_1}, \ldots, c_{\Delta_k})$ of the grid. As there are polynomially many grids  (there are~${n \choose k}$ hyperplanes, with each having at most $(\alpha+1)^k$ grids), and for each grid, checking is done in \sig{NP} (due to SAT problem being~\sig{NP}), the overall process is in~\sig{NP}.  

	\item For the special case where $\mathcal{CS}=\emptyset$, the ``occupation checking'' step mentioned previously is not required. As there are polynomially many grids  (there are~${n \choose k}$ hyperplanes, with each having at most $(\alpha+1)^k$ grids), the overall process is in~\sig{P}.  
	\qed 	
\end{itemize}

\Comment{

\begin{Proposition}
	$\sig{cov}^{1}_{\mathcal{C}}(\mathcal{D}) \geq \sig{cov}^{2}_{\mathcal{C}}(\mathcal{D}) \geq \ldots \geq
	\sig{cov}^{n}_{\mathcal{C}}(\mathcal{D}) = \sig{cov}_{\mathcal{C}}(\mathcal{D})$.
\end{Proposition}

\proof When $k=n$, the projection operator does not change $\sig{sat}(\mathcal{CS})$ and $\sig{to-set}$ operator is not effective as $\sig{sat}(\mathcal{CS})$ is already a set, not a multi-set. Thus the denominator part of Definition~\ref{def.categorization.coverage} and~\ref{def.k.projection} are the same. Similar argument holds for the numerator part. Thus the definition of $	\sig{cov}^{n}_{\mathcal{C}}(\mathcal{D})$ can be rewritten as $\sig{cov}_{\mathcal{C}}(\mathcal{D})$. 

}

\section{Fulfilling $k$-projection Coverage}

As a given data set may not fulfill full $k$-projection coverage, one needs to generate additional data points to increase coverage. By assuming that there exists a \emph{data generator function} $\mathcal{G}$ which can, from any discrete categorization point $c \in \{\sig{0}, \ldots, \alpha\}^n$, creates a new data point $\mathcal{G}(c)$ in $\mathcal{DS}$ such that $ \mathcal{C}(\mathcal{G}(c))= c$ and $\mathcal{G}(c) \not\in \mathcal{D}$ (e.g., for image generation,~$\mathcal{G}$ can be realized using techniques such as conditional-GAN~\cite{mirza2014conditional} to synthesize an image following the specified criterion, or using manually synthesized videos), generating data points to increase coverage amounts to the problem of finding additional discrete categorization points. 

\begin{Definition}[Efficiently increasing $k$-projection coverage]
	Given a data set~$\mathcal{D}$, categorization $\mathcal{C}$ and generator $\mathcal{G}$, the problem of increasing $k$-projection coverage refers to the problem of finding a minimum sized set $\varTheta \subseteq  \{\sig{0}, \ldots, \alpha\}^n $, such that 
	$\emph{\sig{cov}}^{k}_{\mathcal{C}}(\mathcal{D} \cup \{\mathcal{G}(c) : c \in \varTheta \})= 1$.	
\end{Definition}

\paragraph{\textbf{(Book-keeping $k$-projection for a given data set)}}

For $\Delta=\{\Delta_1, \ldots, \Delta_k\}$, we use $\boxed{C_{\Delta_1}\ldots C_{\Delta_k}}$ to represent the data structure for book-keeping the covered items, and use subscript "$_{\countappear{\gamma}}$"  to indicate that certain categorization has been covered $\gamma$ times by the existing data set.

\vspace{1mm}
\noindent{\textbf{(Example 3)}} Consider the following three discrete categorization points
$\{(\sig{0},\sig{0},\sig{1},\sig{1}),\\
 (\sig{1},\sig{0},\sig{0},\sig{0}), (\sig{1},\sig{0},\sig{0},\sig{1})\}$ under $\alpha=\sig{1}$. Results of applying $1$-projection and $2$-projection are book-kept in Equation~\ref{eq.1.projection} and~\ref{eq.2.projection} respectively.
	
\vspace{-3mm}
	\begin{equation}~\label{eq.1.projection}
	\boxed{C_1} = \{\sig{0}_{\countappear{1}}, \sig{1}_{\countappear{2}}\}, \boxed{C_2} =\{ \sig{0}_{\countappear{3}}, \sig{1}_{\countappear{0}}\}, \boxed{C_3} = \{ \sig{0}_{\countappear{2}}, \sig{1}_{\countappear{1}}\},
	\boxed{C_4} = \{\sig{0}_{\countappear{1}}, \sig{1}_{\countappear{2}}\}	
	\end{equation}

\vspace{-5mm}
	\begin{multline}~\label{eq.2.projection}
	\boxed{C_1C_2} = \{\sig{00}_{\countappear{1}}, \sig{01}_{\countappear{0}}, \sig{10}_{\countappear{2}}, \sig{11}_{\countappear{0}}\} \;
\boxed{C_1C_3} = \{\sig{00}_{\countappear{0}}, \sig{01}_{\countappear{1}}, \sig{10}_{\countappear{2}}, \sig{11}_{\countappear{0}}\} \\
\boxed{C_1C_4} = \{\sig{00}_{\countappear{0}}, \sig{01}_{\countappear{1}},  \sig{10}_{\countappear{1}}, \sig{11}_{\countappear{1}}\}\;
\boxed{C_2C_3} = \{\sig{00}_{\countappear{2}}, \sig{01}_{\countappear{1}}, \sig{10}_{\countappear{0}}, \sig{11}_{\countappear{0}}\} \\
\boxed{C_2C_4} = \{\sig{00}_{\countappear{1}}, \sig{01}_{\countappear{2}}, \sig{10}_{\countappear{0}}, \sig{11}_{\countappear{0}}\}\;
\boxed{C_3C_4} = \{\sig{00}_{\countappear{1}}, \sig{01}_{\countappear{1}}, \sig{10}_{\countappear{0}}, \sig{11}_{\countappear{1}}\}
\end{multline}


\paragraph{\textbf{(Full $k$-projection coverage under $\mathcal{CS} = \emptyset$)}}

To achieve $k$-projection coverage under $\mathcal{CS} = \emptyset$, in the worst case, one can always generate ${n \choose k}(\alpha+1)^k\beta^k$ discrete categorization points for $|\varTheta|$ in polynomial time. Precisely, to complete coverage on a particular projection $\Delta$, simply enumerate all possible assignments (a total of $(\alpha+1)^k$ assignments, as $k$ is a constant, the process is done in polynomial time) for all $(C_{\Delta_1}, \ldots, C_{\Delta_k})$, and extend them by associating $C_i$, where $i\in \{1, \ldots, n\}\setminus \Delta$, with arbitrary value within $\{\sig{0}, \ldots, \alpha\}$, and do it for $\beta^k$ times. For example, to increase $2$-projection coverage in Equation~\ref{eq.2.projection}, provided that $W_i(j)=1$, one first completes $\boxed{C_1C_2}$ by adding $\{\sig{01-\;-},\sig{11-\;-}\}$ where ``$\sig{-}$" can be either~$\sig{0}$ or~$\sig{1}$. One further improves $\boxed{C_1C_3}$ using  $\{\sig{0-0-},\sig{1-1-}\}$, and subsequently all others. 

As using $|\varTheta|$ to be ${n \choose k}(\alpha+1)^k\beta^k$ can still create problems when data points are manually generated from discrete categorization points, in the following, we demonstrate important sub-cases with substantially improved bounds over $|\varTheta|$. 

\begin{Proposition}[$1$-projection coverage]
	Finding an additional set of discrete categorization points $\varTheta$ to achieve $1$-projection coverage, with minimum size and under the condition of $\mathcal{CS} = \emptyset$, can be solved in time $\mathcal{O}(\alpha^2 \beta n^2)$, with $|\varTheta|$ being bounded by $(\alpha+1)\beta$. 
\end{Proposition}

\proof We present an algorithm (Algo.~\ref{algo.1.projection}) that allows generating minimum discrete categorization points for full $1$-projection coverage. Recall for $1$-projection, our starting point is  $\boxed{C_{\Delta_1}}, \ldots, \boxed{C_{\Delta_n}}$ with each $\boxed{C_{\Delta_i}}$ recording the number of appearances for element $j \in \{\sig{0}, \ldots, \alpha\}$. We use $\boxed{C_{\Delta_i}}_{[j]}$ to denote the number of appearances for element~$j$ in~$\boxed{C_{\Delta_i}}$. 

\begin{algorithm}[t]
	\KwData{ $\boxed{C_{\Delta_1}}, \ldots, \boxed{C_{\Delta_n}}$ of a given data set, and weight function $\mathcal{W}$}
	\KwResult{The minimum set $\varTheta$ of additional discrete categorization points to guarantee full $1$-projection}
	\While{\sig{true}}{
		\textbf{let} $c:=(\ast,\ldots,\ast)$\;
		\For{$i=1,\ldots,n$}
		{
		\For{$j=\sig{0},\ldots,\alpha$}
		{
			\If{$\boxed{C_{\Delta_i}}_{[j]} < W_i(j)$}{
			replace the $i$-th element of $c$ by value~$j$\;
			$\boxed{C_{\Delta_i}}_{[j]} := \boxed{C_{\Delta_i}}_{[j]} +1$\;
			\textbf{break} /* inner-loop */\;
		}	
		}
		}
		\lIf{$c == (\ast,\ldots,\ast)$}{
			\textbf{return} $\varTheta$
		}\lElse{
			replace every $\ast$ in $c$ by value $\sig{0}$,	$\varTheta := \varTheta \cup \{c\}$
		}
	}
	\caption{Algorithm for achieving $1$-projection.}

\label{algo.1.projection}
\end{algorithm}

In Algo.~\ref{algo.1.projection}, for every projection~$i$, the inner loop picks a value~$j$ whose appearance in~$\boxed{C_{\Delta_i}}$ is lower than $W_i(j)$ (line 5-9). 
If no value is picked for some projection $i$, then the algorithm just replaces $\ast$ by $\sig{0}$, before adding it to the set~$\varTheta$ used to increase coverage (line 13). If after the iteration, $c$ remains to be $(\ast,\ldots,\ast)$, then we have achieved full $1$-projection coverage and the program exits (line 12).
The algorithm guarantees to return a set fulling full $1$-projection with minimum size, due to the observation that each categorization is independent, so the algorithm stops so long as the categorization which misses most elements is completed. In the worst case,  if projection~$i$ started without any data, after $(\alpha+1)\beta$ iterations, it should have reached a state where it no longer requires additional discrete characterization points. Thus, $|\varTheta|$ is guaranteed to be bounded by $(\alpha+1)\beta$. 

Consider the example in Eq.~\ref{eq.1.projection}. When $W_i(j)=1$, the above algorithm reports that only one additional discrete categorization point $(\sig{0}, \sig{1}, \sig{0}, \sig{0})$ is needed to satisfy full $1$-projection. 

\qed

On the other hand, efficiently increasing $3$-projection coverage, even under the condition of $\mathcal{CS} = \emptyset$, is hard. 

\begin{Proposition}[Hardness of maximally increasing $3$-projection coverage, when $\mathcal{CS} = \emptyset$]
	Checking whether there exists one discrete categorization point to increase $3$-projection coverage from existing value $\chi$ to value $\chi'$, under the condition where $\odot$ is scalar multiplication, is \emph{\sig{NP}}-hard. 
\end{Proposition}

\proof (Sketch) The hardness result is via a reduction from 3-SAT satisfiability, where we assume that each clause has exactly three variables. This problem is known to be $\sig{NP}$-complete. We consider the case where $\alpha = \sig{2}$ and $\beta = 1$, i.e., each categorization function creates values in $\{\sig{0}, \sig{1}, \sig {2}\}$.  Given a 3-SAT formula $\phi_{3SAT}$ with $\delta$ clauses, with each literal within the set of variables being~$\{C_1, C_2, \ldots, C_n\}$,  we perform the following construction. 
\begin{itemize}
	\item Set the weight of categorization such that $W_i(\sig{0})=W_i(\sig{1})=1$ and $W_i(\sig{2})=0$.
	\item For each clause such as $(C_x \vee \neg C_y \vee C_z)$, we create a discrete categorization point by setting $C_x = \sig{0}$, $C_y = \sig{1}$, $C_z = \sig{0}$ (i.e., the corresponding assignment makes the clause \sig{false}) and by setting remaining $C_i$ to be $\sig{2}$. Therefore, the process creates a total of $\delta$ discrete categorization points and can be done in polynomial time.
	
	\item Subsequently, prepare the data structure and record the result of $3$-projection for the above created discrete categorization points. As there are at most ${n \choose 3}$ boxes of form  $\boxed{C_x C_y C_z}$, with each box having $|\{\sig{0},\sig{1},\sig{2}\}|^3=27$ items, the construction can be conducted in polynomial time.

	\item One can subsequently compute the $3$-projection coverage. Notice that due to the construction of $W_i(\sig{2})=0$, all projected elements that contain value~$\sig{2}$ should not be counted. The computed denominator should be~${n \choose 3}(2)^3$ rather than~${n \choose 3}(3)^3$ also due to $W_i(\sig{2})=0$. 

\end{itemize}

Then the $\phi_{3SAT}$ problem has a satisfying instance \emph{iff} there exists a discrete categorization point which increases the $3$-projection coverage from  $\frac{a}{{n \choose 3}(2)^3}$ to value $\frac{a+{n \choose 3}}{{n \choose 3}(2)^3}$. 

\begin{itemize}
	\item ($\Rightarrow$) If $\phi_{3SAT}$ has a satisfying instance, create a discrete categorization point where $C_i = \sig{0}$ ($C_i = \sig{1}$) if the satisfying assignment of  $\phi_{3SAT}$, $C_i$ equals $\sig{false}$ ($\sig{true}$). The created discrete categorization point, when being projected, will
	\begin{itemize}
		\item 	not occupy the already occupied space (recall that overlapping with existing items in each box implies that the corresponding clause can not be satisfied), and
		\item not occupy a grid having $C_i = \sig{2}$ (as the assignment only makes~$C_i$ to be \sig{0} or \sig{1}), making the point being added truly help in increasing the numerator of the computed coverage.
	\end{itemize}
Overall, each projection will increase value by~$1$, and therefore, the $3$-projection coverage increases from  $\frac{a}{{n \choose 3}(2)^3}$ to value $\frac{a+{n \choose 3}}{{n \choose 3}(2)^3}$.  
	
	\item ($\Leftarrow$) Conversely, if there exists one discrete categorization point to increase coverage by ${n \choose 3}$, due to the fact that we only have one point and there are ${n \choose 3}$ projections, it needs to increase in \emph{each} box representing $3$-projection, without being overlapped with existing items in that box and without having value $\sig{2}$ being used. One can subsequently use the value of the discrete categorization point to create a satisfying assignment.  \qed
\end{itemize}

In the following, we present an algorithm which encodes the problem of finding a discrete categorization point with maximum coverage increase to a 0-1 integer programming problem. Stated in  Algo.~\ref{algo.encoding}, line~1 prepares variables and constraints to be placed in the 0-1 programming framework. For each categorization $C_i$, for each possible value we create an 0-1 variable $\sig{var}_{[C_i=j]}$ (line 3-5), such that $\sig{var}_{[C_i=j]}=1$ iff the newly introduced discrete categorization point has~$C_i$ using value~$j$. As the algorithm proceeds by only generating one discrete categorization point, only one of them can be true, which is reflected in the constraint  $\sum_{j=0}^{\alpha}\sig{var}_{[ C_i = j]} = 1$ in line~6. 

Then starting from line~8, the algorithm checks if a particular projected value  still allows improvement $\boxed{C_{\Delta_1}\ldots C_{\Delta_k}}_{[v_{\Delta_1}\ldots v_{\Delta_k}]} < W_{\Delta_1}(v_{\Delta_1})\odot \ldots \odot W_{\Delta_k}(v_{\Delta_k})$. If so, then create a variable $\sig{occ}_{[C_{\Delta_1}=v_{\Delta_1}, \ldots, C_{\Delta_k}=v_{\Delta_k}]}$ (line~10) such that it is set to~$1$ iff the newly introduced discrete categorization point will occupy this grid when being projected. As our goal is to maximally increase $k$-projection coverage, $\sig{occ}_{[C_{\Delta_1}=v_{\Delta_1}, \ldots, C_{\Delta_k}=v_{\Delta_k}]}$ is introduced in the objective function (line~11 and~16) where the sum of all variables is the objective to be maximized. 
Note that $\sig{occ}_{[C_{\Delta_1}=v_{\Delta_1}, \ldots, C_{\Delta_k}=v_{\Delta_k}]}$ is set to $1$ iff the newly introduced discrete categorization point guarantees that $C_{\Delta_1}=v_{\Delta_1} \wedge \ldots \wedge C_{\Delta_k}=v_{\Delta_k}$. For this purpose, line~12 applies a standard encoding tactic in 0-1 integer programming to encode such a condition - If $\sig{var}_{[C_{\Delta_1}=v_{\Delta_1}]}= \ldots = \sig{var}_{[C_{\Delta_k}=v_{\Delta_k}]}=1$, then $\sig{var}_{[C_{\Delta_1}=v_{\Delta_1}]} + \ldots + \sig{var}_{[C_{\Delta_k}=v_{\Delta_k}]} =k$. Thus 
$\sig{occ}_{[C_{\Delta_1}=v_{\Delta_1}, \ldots, C_{\Delta_k}=v_{\Delta_k}]}$ will be set to~$1$ to enforce satisfaction of the right-hand inequality of the constraint. Contrarily, if any of $\sig{var}_{[C_{\Delta_j}=v_{\Delta_j}]}$, where $j\in \{1,\ldots, k\}$ has value $0$, then $\sig{occ}_{[C_{\Delta_1}=v_{\Delta_1}, \ldots, C_{\Delta_k}=v_{\Delta_k}]}$ needs to set to~$0$, in order to enforce the left-hand inequality of the constraint. Consider the example in Eq.~\ref{eq.2.projection}, where one has 	$\boxed{C_1C_2} = \{\sig{00}_{\countappear{1}}, \sig{01}_{\countappear{0}}, \sig{10}_{\countappear{2}}, \sig{11}_{\countappear{0}}\}$. For improving $\sig{01}_{\countappear{0}}$, line~12 generates the following constraint $0 \leq \sig{var}_{[C_1=\sig{0}]} + \sig{var}_{[C_2=\sig{1}]} - 2\;\sig{occ}_{[C_1=\sig{0}, C_2=\sig{1}]} \leq 1$.

Line~14 will be triggered when no improvement can be made by every check of line~9, meaning that the system has already achieved full $k$-projection coverage. Lastly, apply 0-1 integer programming where one translates variable $\sig{var}_{[C_i = v_i]} $ having  value~$1$ by assigning $C_i$ to $v_i$ in the newly generated discrete categorization point (line~17,~18).  

Here we omit technical details, but Algo.~\ref{algo.encoding} can easily be extended to constrained cases by adding $\mathcal{CS}$ to the list of constraints.

\begin{algorithm}[t]

	\KwData{ The set $\{\boxed{C_{\Delta_1}\ldots C_{\Delta_k}}\}$ of the current $k$-projection records, and weight function $\mathcal{W}$}
	\KwResult{One discrete categorization point $(c_1, \ldots, c_n)$ which maximally increase coverage, or $\sig{null}$ if current records have achieved full coverage.}
	
	\textbf{let} $\sig{var}_{0/1} := \emptyset$, $\sig{constraints}_{0/1} := \emptyset$, $\sig{objvar}_{0/1} := \emptyset$,\;
	\ForAll{$C_i$, $i \in \{1, \ldots, n\}$}{
	\ForAll{$j \in \{0, \ldots, \alpha\}$}{
			$\sig{var}_{0/1} := \sig{var}_{0/1} \cup 
			\{\sig{var}_{[ C_i = j]}\}$\;	
	}
$\sig{constraints}_{0/1} := \sig{constraints}_{0/1} \cup \{\sum_{j=0}^{\alpha}\sig{var}_{[ C_i = j]} = 1 \}$\;

	}
	 
	\ForAll{$\boxed{C_{\Delta_1}\ldots C_{\Delta_k}}$}{
		\If{$\boxed{C_{\Delta_1}\ldots C_{\Delta_k}}_{[v_{\Delta_1}\ldots v_{\Delta_k}]} < W_{\Delta_1}(v_{\Delta_1})\odot \ldots \odot W_{\Delta_k}(v_{\Delta_k})$}
		{
		$\sig{var}_{0/1} := \sig{var}_{0/1} \cup \{
		\sig{occ}_{[C_{\Delta_1}=v_{\Delta_1}, \ldots, C_{\Delta_k}=v_{\Delta_k}]}
		\} $\;
		
			$\sig{objvar}_{0/1} := \sig{objvar}_{0/1} \cup \{\sig{occ}_{[C_{\Delta_1}=v_{\Delta_1}, \ldots, C_{\Delta_k}=v_{\Delta_k}]}\}$\;	
		$\sig{constraints}_{0/1} := \sig{constraints}_{0/1} \cup\{ 0 \leq \sig{var}_{[C_{\Delta_1}=v_{\Delta_1}]} + \ldots + \sig{var}_{[C_{\Delta_k}=v_{\Delta_k}]} - k \; \sig{occ}_{[C_{\Delta_1}=v_{\Delta_1}, \ldots, C_{\Delta_k}=v_{\Delta_k}]} \leq k-1 \}$\;
	}
	\lIf{$\emph{\sig{objvar}} = \emptyset$}{\textbf{return} \sig{null}}
	\Else{
		\textbf{let} $\sig{obj} :=  \sum_{\sig{var} \in \sig{objvar} } \sig{var}$\;
		
		\textbf{let} $\sig{assignment} := \sig{0/1-programming}_{\{\sig{var}_{0/1}\}}(\textbf{maximize} \;\;\sig{obj} \;\; \textbf{subject-to} \;\;\sig{constraint}_{0/1})$\; 
		\textbf{return} $(v_1, \ldots, v_n)$ where in \sig{assignment} $\sig{var}_{[C_i = v_i]} $ is assigned to $1$\;
	}
	}

		\caption{Finding a discrete categorization point which maximally increases $k$-projection coverage, via an encoding to 0-1 integer programming.}
\label{algo.encoding}
\end{algorithm}

\section{Implementation and Evaluation}

\begin{figure}[t]
			\includegraphics[width=\textwidth]{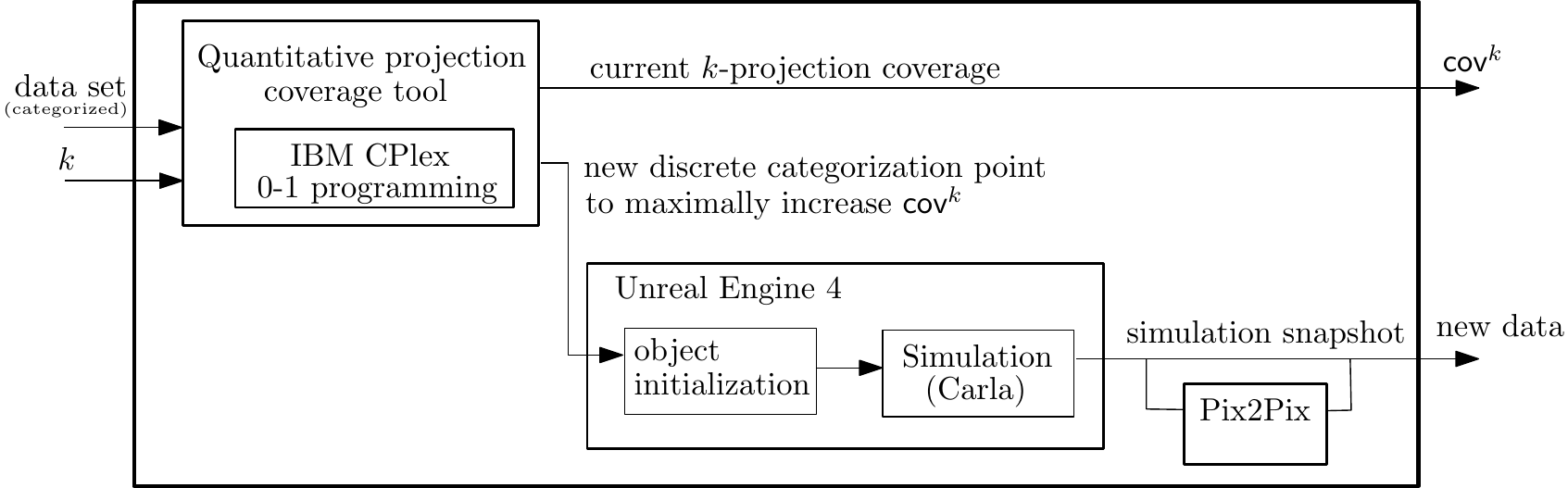}
			\caption{Workflow in the developed prototype for quantitative projection coverage and generation of new synthetic data.}%
			\label{fig:implementation}
\end{figure}

\begin{figure}[t]
\vspace{-5mm}
	\includegraphics[width=1.05\textwidth]{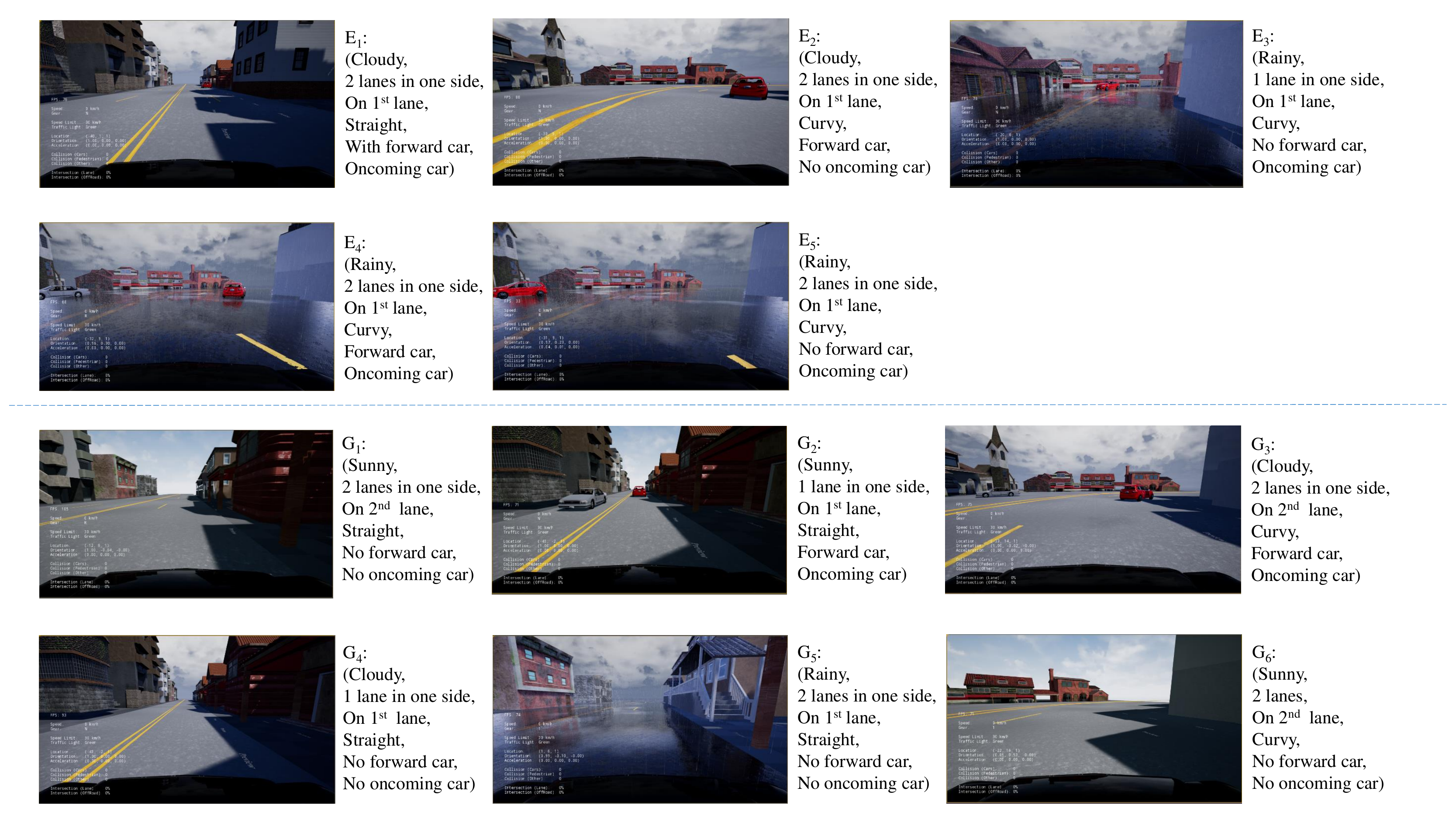}
	\caption{Existing data points (E$_1$ to E$_5$), and the automatically generated data points (G$_1$ to G$_6$) to achieve full coverage.}%
	\label{fig:Image_of_DataPoints}
\end{figure}

We have implemented above mentioned techniques as a workbench to support  training vision-based perception units for autonomous driving. The internal workflow of our developed tool is summarized in  Fig.~\ref{fig:implementation}. It takes existing labelled/categorized data set and the user-specified~$k$ value as input, computes  $k$-projection coverage, and finds a new discrete categorization point which can increase the coverage most significantly. For the underlying 0-1 programming solving, we use IBM CPLEX Optimization Studio\footnote{IBM CPLEX Optimization Studio: \url{https://www.ibm.com/analytics/data-science/prescriptive-analytics/cplex-optimizer}}.

To convert the generated discrete categorization points to real images, we have further implemented a C++ plugin over the Carla simulator\footnote{Carla Simulator: \url{http://carla.org/}}, an open-source simulator for autonomous driving based on Unreal Engine 4\footnote{Unreal Engine 4: \url{https://www.unrealengine.com/}}.
The plugin reads the scenario from the discrete categorization point and configures the ground truth, the weather, and the camera angle accordingly. Then the plugin starts the simulation and takes a snapshot using the camera mounted on the simulated vehicle. The camera can either return synthetic images (e.g., images in Fig.~\ref{fig:Image_of_DataPoints}) or images with segmentation information, where for the latter one, we further generate close-to-real image via applying conditional GAN framework Pix2Pix from NVIDIA\footnote{\url{https://github.com/NVIDIA/pix2pixHD}}. Due to space limits, here we detail a small example by choosing the following operating conditions of autonomous vehicles as our categories.

\vspace{-1mm}

\begin{itemize}
	\item Weather = $\{\sig{Sunny}, \sig{Cloudy}, \sig{Rainy}\}$
	\item Lane orientation = $\{\sig{Straight}, \sig{Curvy}\}$
	\item Total number of lanes (one side) = $\{1, 2\}$
	\item Current driving lane = $\{1, 2\}$
	\item Forward vehicle existing = $\{\sig{true}, \sig{false}\}$
	\item  Oncoming vehicle existing = $\{\sig{true}, \sig{false}\}$
\end{itemize}

\vspace{-1mm}

\begin{figure}[t]
	\includegraphics[width=0.6\textwidth]{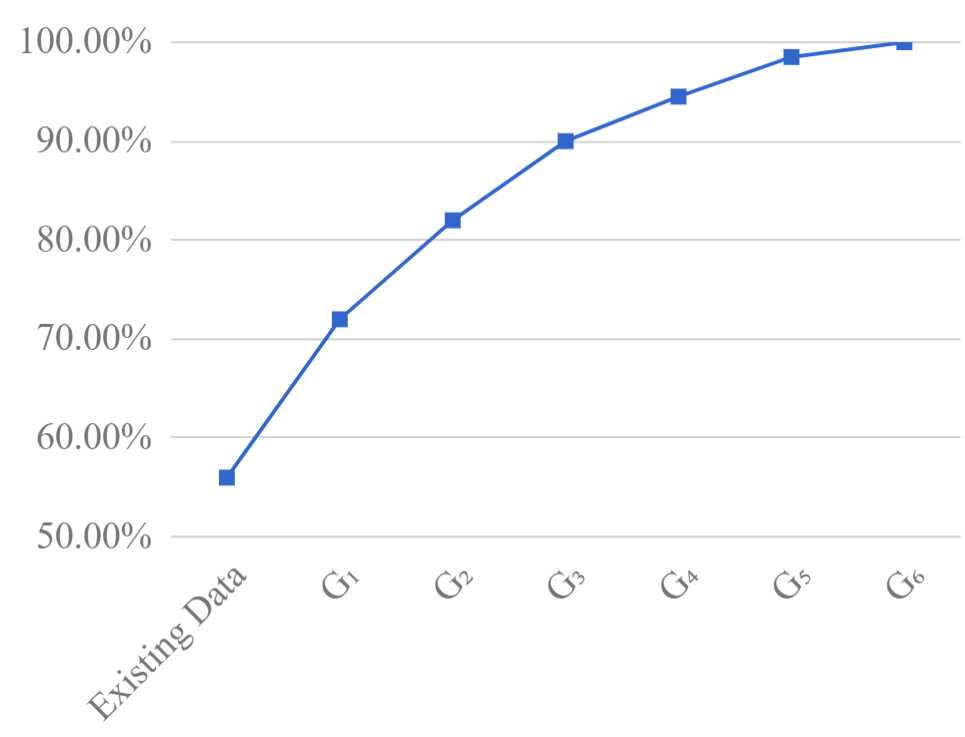}
	\vspace{-2mm}
	\caption{Change of $2$-projection coverage due to newly generated data.}%
	\label{fig:Growth}
\end{figure}

We used our test case generator to generate new data points to achieve full $2$-projection coverage (with $W_i(j)=1$) starting with a small set of randomly captured data points (Fig.~\ref{fig:Image_of_DataPoints}, images E$_1$ to E$_5$). Images G$_1$ to G$_6$ are synthesized in sequence until full $2$-projection coverage is achieved. The coverage condition of each 2-projection plane is shown in Table~\ref{tab_CoverageTables}. Note that there exists one entry in the sub-table (f) which is not coverable  (labelled as ``X''), as there is a constraint stating that \textit{if there exists only~$1$ lane, it is impossible for the vehicle to drive on the $2^{nd}$ lane}. Fig.~\ref{fig:Growth} demonstrates the growth of $2$-projection coverage when gradually introducing images~G$_1$ to~G$_6$.

\begin{table}[t]

		\begin{subtable}[t]{.32\linewidth}
		\centering
		\begin{tabular}{|c|c|c|}
			\hline
			&  1 Lane & 2 Lanes      \\ \hline
			Sunny  &  G$_2$ & G$_1$  \\ \hline
			Cloudy &  G$_4$ & G$_3$  \\ \hline
			Rainy  &  E$_3$ & E$_4$  \\ \hline
		\end{tabular}
		\caption{Weather \& $\#$Lanes}
	\end{subtable}
	\begin{subtable}[t]{.32\linewidth}
		\centering
		\begin{tabular}{|c|c|c|}
			\hline
			&  $1^{st}$ Lane  & $2^{nd}$ Lane      \\ \hline
			Sunny  &  G$_2$ & G$_1$  \\ \hline
			Cloudy &  G$_4$ & G$_3$  \\ \hline
			Rainy  &  E$_5$ & G$_5$  \\ \hline
		\end{tabular}
		\caption{Weather \& Current Lane}
	\end{subtable}
	\begin{subtable}[t]{.32\linewidth}
		\centering
		\begin{tabular}{|c|c|c|}
			\hline
			&  Straight  & Curvy    \\ \hline
			Sunny  &  G$_1$     & G$_6$    \\ \hline
			Cloudy &  G$_4$     & G$_3$    \\ \hline
			Rainy  &  G$_5$     & E$_5$    \\ \hline
		\end{tabular}
		\caption{Weather \& Lane Curve}
	\end{subtable}
	\begin{subtable}[t]{.32\linewidth}
		\centering
		\begin{tabular}{|c|c|c|}
			\hline
			&  No FC     &   FC    \\ \hline
			Sunny  &  G$_1$     & G$_2$    \\ \hline
			Cloudy &  G$_4$     & G$_3$    \\ \hline
			Rainy  &  G$_5$     & E$_4$    \\ \hline
		\end{tabular}
		\caption{Weather \& Forward Car}
	\end{subtable}
	\begin{subtable}[t]{.32\linewidth}
		\centering
		\begin{tabular}{|c|c|c|}
			\hline
			&  No OC     &   OC    \\ \hline
			Sunny  &  G$_1$     & G$_2$    \\ \hline
			Cloudy &  G$_4$     & G$_3$    \\ \hline
			Rainy  &  G$_5$     & E$_4$    \\ \hline
		\end{tabular}
		\caption{Weather \& Oncoming Car}
	\end{subtable}
	\begin{subtable}[t]{.32\linewidth}
		\centering
		\begin{tabular}{|c|c|c|}
			\hline
			&  $1^{st}$ Lane & $2^{nd}$ Lane    \\ \hline
			1 Lane   &  E$_3$     & X  \\ \hline
			2 Lanes  &  E$_2$     & G$_1$    \\ \hline
		\end{tabular}
		\caption{$\#$Lanes \& Current Lane}
	\end{subtable}
	\begin{subtable}[t]{.32\linewidth}
		\centering
		\begin{tabular}{|c|c|c|}
			\hline
			&  Straight  & Curvy    \\ \hline
			1 Lane   &  G$_2$     & E$_3$   \\ \hline
			2 Lanes  &  E$_1$     & E$_2$    \\ \hline
		\end{tabular}
		\caption{$\#$Lanes \&  Lane Curve}
	\end{subtable}
	\begin{subtable}[t]{.32\linewidth}
		\centering
		\begin{tabular}{|c|c|c|}
			\hline
			&  No FC  & FC    \\ \hline
			1 Lane   &  E$_3$  & G$_2$   \\ \hline
			2 Lanes  &  G$_1$  & G$_3$    \\ \hline
		\end{tabular}
		\caption{$\#$Lanes \& Forward Car}
	\end{subtable}
	\begin{subtable}[t]{.32\linewidth}
		\centering
		\begin{tabular}{|c|c|c|}
			\hline
			&  No OC  & OC    \\ \hline
			1 Lane   &  G$_4$  & E$_3$   \\ \hline
			2 Lanes  &  E$_1$  & E$_4$    \\ \hline
		\end{tabular}
		\caption{$\#$Lanes \& Oncoming Car}
	\end{subtable}
	\begin{subtable}[t]{.32\linewidth}
		\centering
		\begin{tabular}{|c|c|c|}
			\hline
			&  Straight  & Curvy    \\ \hline
			$1^{st}$ Lane   &  E$_1$  & E$_2$   \\ \hline
			$2^{nd}$ Lane   &  G$_1$  & G$_3$    \\ \hline
		\end{tabular}
		\caption{Current Lane \& Lane Curve}
	\end{subtable}
	\begin{subtable}[t]{.32\linewidth}
		\centering
		\begin{tabular}{|c|c|c|}
			\hline
			&  No FC  & FC    \\ \hline
			$1^{st}$ Lane   &  E$_3$  & E$_1$   \\ \hline
			$2^{nd}$ Lane   &  G$_1$  & G$_3$    \\ \hline
		\end{tabular}
		\caption{Current Lane \& Forward Car}
	\end{subtable}
	\begin{subtable}[t]{.32\linewidth}
		\centering
		\begin{tabular}{|c|c|c|}
			\hline
			&  No OC  & OC    \\ \hline
			$1^{st}$ Lane   &  E$_1$  & E$_3$   \\ \hline
			$2^{nd}$ Lane   &  G$_1$  & G$_3$    \\ \hline
		\end{tabular}
		\caption{Current Lane \& Oncoming Car}
	\end{subtable}
	\begin{subtable}[t]{.32\linewidth}
		\centering
		\begin{tabular}{|c|c|c|}
			\hline
			&  No FC  & FC    \\ \hline
			Straight   &  G$_1$  & E$_1$   \\ \hline
			Curvy      &  E$_3$  & E$_2$    \\ \hline
		\end{tabular}
		\caption{Lane Curve \& Forward Car}
	\end{subtable}
	\begin{subtable}[t]{.32\linewidth}
		\centering
		\begin{tabular}{|c|c|c|}
			\hline
			&  No OC  & OC    \\ \hline
			Straight   &  E$_1$  & G$_2$   \\ \hline
			Curvy      &  E$_4$  & E$_2$    \\ \hline
		\end{tabular}
		\caption{Lane Curve \& Oncoming Car}
	\end{subtable}
	\begin{subtable}[t]{.32\linewidth}
		\centering
		\begin{tabular}{|c|c|c|}
			\hline
			&  No OC  & OC    \\ \hline
			No FC   &  G$_1$  & E$_3$   \\ \hline
			FC      &  E$_2$  & E$_4$    \\ \hline
		\end{tabular}
		\caption{Forward Car \& Oncoming Car}
	\end{subtable}
		
	\caption{$2$-projection coverage tables of the final data set}\label{tab_CoverageTables}
\end{table}

\section{Concluding Remarks}

In this paper, we presented quantitative $k$-projection coverage as a method to systematically evaluate the quality of data for systems engineered using machine learning approaches. Our prototype implementation is used to compute coverage and synthesize additional images for engineering a vision-based perception unit for automated driving. The proposed metric can further be served as basis to refine other classical metrics such as MTBF or availability which is based on statistical measurement. 

Currently, our metric is to take more data points for  important (higher weight) scenarios.  
For larger~$k$ values, achieving full  projection coverage may not be feasible, so one extension is to adapt the objective function of Algo.~\ref{algo.encoding} such that the generation process favors discrete categorization points with higher weights when being projected. Another direction is to improve the encoding of Algo.~\ref{algo.encoding} such that the algorithm can return multiple discrete categorization points instead of one. Yet another direction is to further associate temporal behaviors to 
 categorization and the associated categorization constraints, when the data space represents a sequence of images.

\vspace{-2mm}
\bibliographystyle{abbrv}


\begin{thebibliography}{10}
	
	\bibitem{DBLP:journals/corr/BojarskiTDFFGJM16}
	M.~Bojarski, D.~D. Testa, D.~Dworakowski, B.~Firner, B.~Flepp, P.~Goyal, L.~D.
	Jackel, M.~Monfort, U.~Muller, J.~Zhang, X.~Zhang, J.~Zhao, and K.~Zieba.
	\newblock End to end learning for self-driving cars.
	\newblock {\em CoRR}, abs/1604.07316, 2016.
	
	\bibitem{carlini2017towards}
	N.~Carlini and D.~Wagner.
	\newblock Towards evaluating the robustness of neural networks.
	\newblock In {\em Security and Privacy (SP), 2017 IEEE Symposium on}, pages
	39--57. IEEE, 2017.
	
	\bibitem{chen2015deepdriving}
	C.~Chen, A.~Seff, A.~Kornhauser, and J.~Xiao.
	\newblock Deepdriving: Learning affordance for direct perception in autonomous
	driving.
	\newblock In {\em Proceedings of the IEEE International Conference on Computer
		Vision}, pages 2722--2730, 2015.
	
	\bibitem{cheng2018neural}
	C.-H. Cheng, F.~Diehl, G.~Hinz, Y.~Hamza, G.~N{\"u}hrenberg, M.~Rickert,
	H.~Ruess, and M.~Truong-Le.
	\newblock Neural networks for safety-critical applications—challenges,
	experiments and perspectives.
	\newblock In {\em Design, Automation \& Test in Europe Conference \& Exhibition
		(DATE), 2018}, pages 1005--1006. IEEE, 2018.
	
	\bibitem{cheng2017maximum}
	C.-H. Cheng, G.~N{\"u}hrenberg, and H.~Ruess.
	\newblock Maximum resilience of artificial neural networks.
	\newblock In {\em International Symposium on Automated Technology for
		Verification and Analysis}, pages 251--268. Springer, 2017.
	
	\bibitem{colbourn2004combinatorial}
	C.~J. Colbourn.
	\newblock Combinatorial aspects of covering arrays.
	\newblock {\em Le Matematiche}, 59(1, 2):125--172, 2004.
	
	\bibitem{ehlers2017formal}
	R.~Ehlers.
	\newblock Formal verification of piece-wise linear feed-forward neural
	networks.
	\newblock {\em arXiv preprint arXiv:1705.01320}, 2017.
	
	\bibitem{DBLP:conf/cav/HuangKWW17}
	X.~Huang, M.~Kwiatkowska, S.~Wang, and M.~Wu.
	\newblock Safety verification of deep neural networks.
	\newblock In {\em Computer Aided Verification - 29th International Conference,
		{CAV} 2017, Heidelberg, Germany, July 24-28, 2017, Proceedings, Part {I}},
	pages 3--29, 2017.
	
	\bibitem{huval2015empirical}
	B.~Huval, T.~Wang, S.~Tandon, J.~Kiske, W.~Song, J.~Pazhayampallil,
	M.~Andriluka, P.~Rajpurkar, T.~Migimatsu, R.~Cheng-Yue, et~al.
	\newblock An empirical evaluation of deep learning on highway driving.
	\newblock {\em arXiv preprint arXiv:1504.01716}, 2015.
	
	\bibitem{DBLP:conf/cav/KatzBDJK17}
	G.~Katz, C.~W. Barrett, D.~L. Dill, K.~Julian, and M.~J. Kochenderfer.
	\newblock Reluplex: An efficient {SMT} solver for verifying deep neural
	networks.
	\newblock In {\em Computer Aided Verification - 29th International Conference,
		{CAV} 2017, Heidelberg, Germany, July 24-28, 2017, Proceedings, Part {I}},
	pages 97--117, 2017.
	
	\bibitem{kolter2017provable}
	J.~Z. Kolter and E.~Wong.
	\newblock Provable defenses against adversarial examples via the convex outer
	adversarial polytope.
	\newblock {\em arXiv preprint arXiv:1711.00851}, 2017.
	
	\bibitem{lawrence2011survey}
	J.~Lawrence, R.~N. Kacker, Y.~Lei, D.~R. Kuhn, and M.~Forbes.
	\newblock A survey of binary covering arrays.
	\newblock {\em the electronic journal of combinatorics}, 18(1):84, 2011.
	
	\bibitem{lei1998parameter}
	Y.~Lei and K.-C. Tai.
	\newblock In-parameter-order: A test generation strategy for pairwise testing.
	\newblock In {\em High-Assurance Systems Engineering Symposium, 1998.
		Proceedings. Third IEEE International}, pages 254--261. IEEE, 1998.
	
	\bibitem{Lenz2017}
	D.~Lenz, F.~Diehl, M.~Troung~Le, and A.~Knoll.
	\newblock Deep neural networks for markovian interactive scene prediction in
	highway scenarios.
	\newblock In {\em IEEE Intelligent Vehicles Symposium (IV) 2017}. IEEE, 2017.
	
	\bibitem{mirza2014conditional}
	M.~Mirza and S.~Osindero.
	\newblock Conditional generative adversarial nets.
	\newblock {\em arXiv preprint arXiv:1411.1784}, 2014.
	
	\bibitem{moosavi2016deepfool}
	S.~M. Moosavi~Dezfooli, A.~Fawzi, and P.~Frossard.
	\newblock Deepfool: a simple and accurate method to fool deep neural networks.
	\newblock In {\em Proceedings of 2016 IEEE Conference on Computer Vision and
		Pattern Recognition (CVPR)}, 2016.
	
	\bibitem{nie2011survey}
	C.~Nie and H.~Leung.
	\newblock A survey of combinatorial testing.
	\newblock {\em ACM Computing Surveys (CSUR)}, 43(2):11, 2011.
	
	\bibitem{seroussi1988vector}
	G.~Seroussi and N.~H. Bshouty.
	\newblock Vector sets for exhaustive testing of logic circuits.
	\newblock {\em IEEE Transactions on Information Theory}, 34(3):513--522, 1988.
	
	\bibitem{simonyan2013deep}
	K.~Simonyan, A.~Vedaldi, and A.~Zisserman.
	\newblock Deep inside convolutional networks: Visualising image classification
	models and saliency maps.
	\newblock {\em arXiv preprint arXiv:1312.6034}, 2013.
	
	\bibitem{sinha2017certifiable}
	A.~Sinha, H.~Namkoong, and J.~Duchi.
	\newblock Certifiable distributional robustness with principled adversarial
	training.
	\newblock {\em arXiv preprint arXiv:1710.10571}, 2017.
	
	\bibitem{sun2017fast}
	L.~Sun, C.~Peng, W.~Zhan, and M.~Tomizuka.
	\newblock A fast integrated planning and control framework for autonomous
	driving via imitation learning.
	\newblock {\em arXiv preprint arXiv:1707.02515}, 2017.
	
	\bibitem{sun2018testing}
	Y.~Sun, X.~Huang, and D.~Kroening.
	\newblock Testing deep neural networks.
	\newblock {\em arXiv preprint arXiv:1803.04792}, 2018.
	
	\bibitem{sun2018concolic}
	Y.~Sun, M.~Wu, W.~Ruan, X.~Huang, M.~Kwiatkowska, and D.~Kroening.
	\newblock Concolic testing for deep neural networks.
	\newblock {\em arXiv preprint arXiv:1805.00089}, 2018.
	
	\bibitem{szegedy2013intriguing}
	C.~Szegedy, W.~Zaremba, I.~Sutskever, J.~Bruna, D.~Erhan, I.~Goodfellow, and
	R.~Fergus.
	\newblock Intriguing properties of neural networks.
	\newblock {\em arXiv preprint arXiv:1312.6199}, 2013.
	
	\bibitem{wang2017high}
	T.-C. Wang, M.-Y. Liu, J.-Y. Zhu, A.~Tao, J.~Kautz, and B.~Catanzaro.
	\newblock High-resolution image synthesis and semantic manipulation with
	conditional gans.
	\newblock {\em arXiv preprint arXiv:1711.11585}, 2017.
	
	\bibitem{weng2018towards}
	T.-W. Weng, H.~Zhang, H.~Chen, Z.~Song, C.-J. Hsieh, D.~Boning, I.~S. Dhillon,
	and L.~Daniel.
	\newblock Towards fast computation of certified robustness for relu networks.
	\newblock {\em arXiv preprint arXiv:1804.09699}, 2018.
	
\end{thebibliography}

\end{document}